# Weaponizing Language Models for Cybersecurity Offensive Operations: Automating Vulnerability Assessment Report Validation; A Review Paper

Abdulrahman S Almuhaidib[1, a)], Azlan Mohd Zain[2, b)], Zalmiyah Zakaria[2, c)], Izyan Izzati Kamsani[2, d)] and Abdulaziz S Almuhaidib[3, e)]

[1]*Networks and Communications Department, College of Computer Science and Information Technology, Imam Abdulrahman Bin Faisal University, P.O. Box 1982, Dammam 31441, Saudi Arabia.*
[2]*2Faculty of Computing, Universiti Teknologi Malaysia, Johor Bahru, Malaysia.*
[3]*Research and Development, Inteli Dexer, Dammam, Saudi Arabia*

[a)] *Corresponding author: asmalmuhaidib@iau.edu.sa*
[b)] *Corresponding author: azlanmz@utm.my*
[c)] *zalmiyah@utm.my*
[d)] izyanizzati@utm.my
[e)] abdulaziz@intelidexer.com

**Abstract.** This, with the ever-increasing sophistication of cyberwar, calls for novel solutions. In this regard, Large Language Models (LLMs) have emerged as a highly promising tool for defensive and offensive cybersecurity-related strategies. While existing literature has focused much on the defensive use of LLMs, when it comes to their offensive utilization, very little has been reported-namely, concerning Vulnerability Assessment (VA) report validation. Consequentially, this paper tries to fill that gap by investigating the capabilities of LLMs in automating and improving the validation process of the report of the VA. From the critical review of the related literature, this paper hereby proposes a new approach to using the LLMs in the automation of the analysis and within the validation process of the report of the VA that could potentially reduce the number of false positives and generally enhance efficiency. These results are promising for LLM automatization for improving validation on reports coming from VA in order to improve accuracy while reducing human effort and security postures. The contribution of this paper provides further evidence about the offensive and defensive LLM capabilities and therefor helps in devising more appropriate cybersecurity strategies and tools accordingly.

## INTRODUCTION

The evolving landscape of cyber warfare demands innovative solutions to counter increasingly sophisticated threats, particularly in the context of escalating attacks on critical infrastructure and personal devices. LLMs, such as Meta's Llama, have emerged as powerful tools with potential applications in both defensive and offensive cybersecurity strategies. While the defensive applications of LLMs have been extensively researched, their offensive potential, particularly in automating and enhancing vulnerability assessment report validation, remains largely unexplored [1]–[3].

This research addresses the critical need for more efficient and accurate VA methods by exploring the untapped potential of LLMs in automating the validation of VA reports. Current VA tools often generate reports with a high number of false positives, making it time-consuming to do manual validation. This paper investigates the capabilities of LLMs in Natural Language Processing (NLP) and cybersecurity to automate the analysis and interpretation of these reports, which could potentially reduce human effort and improve the overall effectiveness of VA. Furthermore, this work aims to demonstrate how LLMs can contribute to the development of more efficient and reliable automated cybersecurity tools and strategies to match the evolving threat landscape. Figure 1 summarizes the structure of the literature review [1], [2], [4].

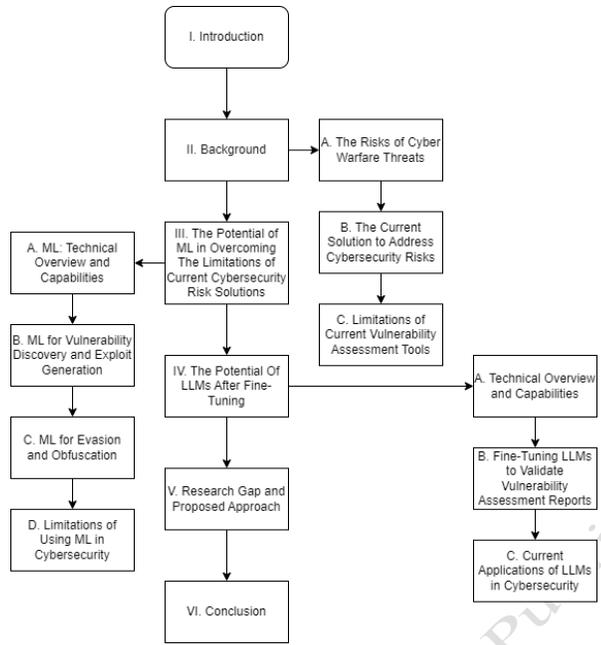

Figure 1. The structure of the paper (Created by the authors)

# BACKGROUND

## The Risks of Cyber Warfare Threats

Offensive cyber-warfare operations are the deliberate cyberspace actions wittingly performed by state or non-state actors with the aim of achieving a number of objectives. Generally, these activities fall into some categories like espionage, sabotage, denial-of-service attacks, and information warfare [1]. Operations of this sort have constantly changed tactics because attackers would employ subtle means of zero-day exploits, malware, and even social engineering to evade detection. Even Artificial Intelligence and Machine Learning (ML) are these days increasingly used in such operations where the attackers use them for automating the attacks, evading detection, and devising newer attack techniques [5].

Such attacks could be catastrophic in several ways: on the personal, organizational, and governmental levels. A pertinent illustration is a sophisticated cyberattack breaching an organization's defenses, resulting in the exfiltration of sensitive information, substantial financial losses, and significant operational disruption. The impact of such attacks extends beyond the disruption to businesses. The integrity of the digital infrastructure would be severely compromised. This hypothetical scenario underscores the critical importance of robust cybersecurity measures. Efficient incident response through a comprehensive risk management approach therefore plays an important role in the mitigation of adverse impact of cybersecurity threats. It is prudent that an organization take drastic measures to secure itself and retain confidence in the stakeholders' eyes by having proper planning in the occurrence of any form of cyber-attack. Preparedness in facing a potential cyber-attack reduces the risk of the damage that otherwise stands at millions of dollars. Figure 2 summarizes the huge financial losses caused by cybersecurity breaches only in the United States [6], [7].

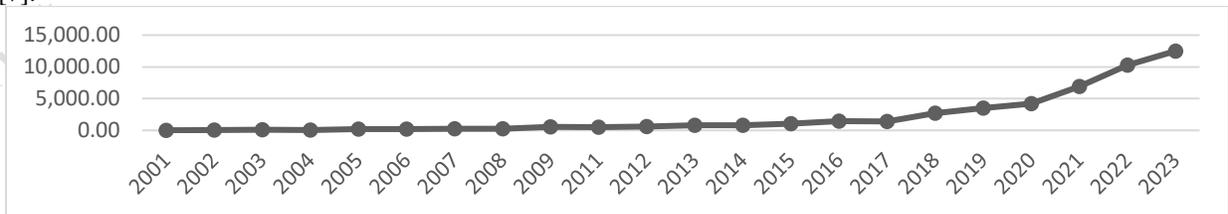

Figure 2. Damages caused by cybercrime in the United States in million dollars (Adopted from[7])

## The Current Solution to Address Cybersecurity Risks

Approaches to defend against such operations were followed in combat against such risks of cybersecurity incidents. In addition, such approaches were put in place to reduce the probability of being exploited by a malicious adversary operation. Each of these approaches has different ways to test them with tests including offensive operations which are simulations of real-life scenarios when it comes to being attacked by a malicious agent to test the resilience of the organization. Another of the most-used tests involves the use of VA tools: software apps intended to scan the network, system, and application for security weaknesses on record regularly [8]. Additionally, technologies such as blockchain have been explored in various domains to enhance data integrity and security by ensuring transparency and preventing tampering [9]. These tools become an important way to find out potential entry points an attacker may detect and help an organization make sure that security risks are dealt with well in advance. VA tools match the configuration and software versions of the target system against a database of known vulnerabilities [10]. They are able to perform active scans, whereby active probes are sent onto the target system in order to get responses that could indicate vulnerabilities. Popular tools in this class include Nessus, OpenVAS, and Qualys Guard. These tools have a host of features including automated scanning, prioritization of vulnerabilities, and reporting.

## Limitations of Current VA Tools

However, several limitations can hinder the effective utilization of VA tools. Such challenges include the quite high number of false positives coming from the tool, showing the presence of a vulnerability where this does not exist in the first place [4]. This leads to wasted hours and resources trying to fix problems that do not exist in reality. On the other hand, the false negatives, when the tool misses the identification of a real vulnerability, are even more dangerous since they leave the system exposed to a possible attack.

The other limitation is that it leaves the actual analysis and prioritization of VA results to human judgment. Most of the tools are voluminous reports, best analyzed by skilled security analysts who help assess the results and pinpoint the most important vulnerabilities to address. That makes the entire process somewhat slow and prone to errors in big and complex environments [11].

# THE POTENTIAL OF ML IN OVERCOMING THE LIMITATIONS OF CURRENT CYBERSECURITY RISK SOLUTIONS

Applications of ML in cybersecurity represent one of the fastest-growing areas in both defensive and offensive use cases. In particular, in this offensive area, ML was applied to perform several tasks, including malware tools to help attackers develop more sophisticated and evasive malware [12]. Beyond cybersecurity, ML has also demonstrated effectiveness in various fields such as medical diagnostics, where it has been used to predict diseases like Parkinson's with high accuracy [13]. Furthermore, in social engineering, ML can personalize phishing attacks and other social engineering tactics, increasing their likelihood of deceiving the target [14]. Another application is automated exploit development. Applying ML algorithms to automate some processes in the development of exploits for known vulnerabilities would speed up the weaponization of those vulnerabilities [15]. For the time being, ML appliances in offensive cybersecurity remain in their early stage, having a huge potential which may significantly change the threat landscape. As ML algorithms become increasingly sophisticated and available, attackers will likely be more keen to leverage them for developing more successful and elusive attack methods.

## ML: Technical Overview and Capabilities

Until now, most research on incorporating ML into cybersecurity has focused on enhancing the systems' defense capabilities. In turn, offensive cyber operations can be greatly empowered by ML, and this is particularly so for penetration testing. Some ML techniques in use cases throughout an offensive cyber-attack are discussed in the next section. This further enhances the automation and efficiency that such an attack can reach [16].

## ML for Vulnerability Discovery and Exploit Generation

ML algorithms have shown promise in automating the discovery of vulnerabilities and the generation of exploits. Supervised learning techniques, such as decision trees [17] and support vector machines [18], can be trained on labeled datasets of known vulnerabilities and exploits to identify patterns and predict potential weaknesses in target systems. These models can then be used to generate exploit code or suggest attack vectors for further investigation. For instance, Valea and Oprisa [17] employed a decision tree algorithm to select the most appropriate exploit based on the target system's operating system, active services, and known Common Vulnerabilities and Exposures (CVEs).

A few of them are unsupervised learning techniques that include clustering and anomaly detection, which can analyze massive volumes of data to find unusual patterns or behaviors that show potential vulnerabilities [12]. They can be effective, especially in uncovering zero-day vulnerabilities that have not yet been discovered by the security community. Clustering algorithms can group similar vulnerabilities, making it easier for an attacker to detect and use them.

Furthermore, reinforcement learning algorithms, such as Q-learning and Deep Q-Networks, can be used to automate the process of penetration testing by learning from experience and adapting to the target environment [5]. These algorithms can explore different attack paths and learn to exploit vulnerabilities more efficiently and effectively. For instance, Goh et al. [19] utilized reinforcement learning to intelligently discover a large number of exploits in a target machine through automated penetration testing. Figure 3 provides a visual representation of the applications of ML for Vulnerability Discovery and Exploit Generation.

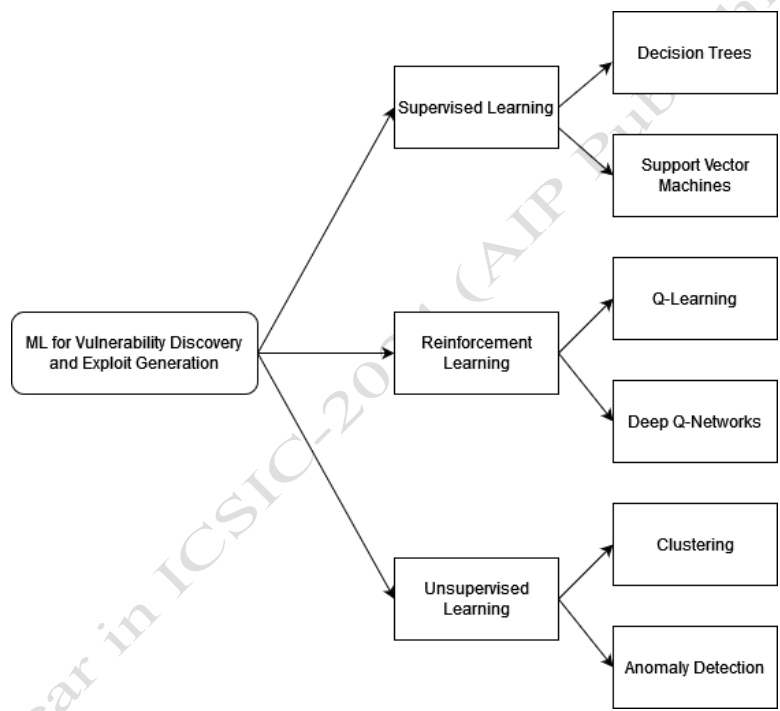

Figure 3. Applications of ML for Vulnerability Discovery and Exploit Generation (Created by the authors)

## ML for Evasion and Obfuscation

ML can be employed for offensive purposes, such as evading detection by security systems and obfuscating malicious code. Adversarial ML techniques can be used to manipulate input data or generate adversarial examples that can bypass Intrusion Detection Systems (IDS) and antivirus software [14]. This can make it difficult for defenders to detect and mitigate attacks, as the malicious activity may appear benign to the security systems. Furthermore, ML can be used to generate polymorphic and metamorphic malware, which can change their code or behavior to evade detection [20]. This makes it more challenging for security tools to identify and mitigate these threats, as the malware constantly changes its signature.

## Limitations of Using ML in Cybersecurity

Even though using ML for offensive cybersecurity operations has high potential, it may still not be the ultimate solution to the problem. Fully automating penetration testing necessitates proficiency in numerous areas, such as natural language processing, code manipulation, network analysis, operating system and tool expertise, and penetration testing methodologies. It is improbable that a single ML model can encompass all these competencies.

However, developing offense cybersecurity ML models is a resource-intensive, complex activity requiring a broad range of expertise in ML and cybersecurity. This involves relevant datasets and computational resources. Development in cyber war is all the more complicated since generally, there is continuous activity wherein the attacker tries to outsmart the defender and vice versa. The major challenge in developing offensive ML models is the need for high-quality training data. Usually, it is very hard to get labeled datasets of real-world attacks, and in most cases, such data are sensitive; hence, supervised learning techniques that learn from labeled data to make predictions are limited by this very fact. Besides, generalization and adaptability will be required. Similarly, models trained with small-sized datasets tend to fail in performing their functions when there is a need for generalization into new or unseen attack scenarios. That is, in this specific domain of cyber warfare, an attacker always renews his tactics and techniques. Therefore, ML models should prove their adaptability against ever-changing threats.

Finally, the interpretability and explainability of ML models are important considerations, especially in security-critical applications. Many ML models, particularly deep learning models, are often considered "black boxes," making it difficult to understand the reasoning behind their decisions. This lack of interpretability can hinder their adoption in offensive cybersecurity, as it may be difficult to trust the model's recommendations or assess the potential impact of its actions. To overcome this limitation, this research proposes the use of LLMs which are a specific application of ML that could handle most of these tasks, and with fine-tuning could potentially achieve full automation.

# THE POTENTIAL OF LLMS AFTER FINE-TUNING

## Technical Overview and Capabilities

LLMs are a class of ML models studied on huge volumes of textual data, including Llama 3. Deep learning methodologies, namely transformer structures, applied to their training enable comprehension of texts remarkably similar to human languages. LLMs have demonstrated proficiency in a diverse array of tasks, including language translation, code processing, network analysis, operating system and tool manipulation, and even penetration testing. Its remarkable ability to analyze and generate human language has led to its diverse application in different fields of study, including cybersecurity [3].

## Current Applications of LLMs in Cybersecurity

The application of LLMs in cybersecurity is a rapidly growing field, with both defensive and offensive use cases being explored. In defensive cybersecurity, LLMs have been used for multiple tasks including vulnerability detection where LLMs analyze code repositories and identify potential security vulnerabilities, aiding in proactive security assessments [2]. LLMs also can process vast amounts of threat intelligence data to identify patterns and trends, helping security analysts stay ahead of emerging threats [4]. Another application is phishing detection by analyzing emails and other communications to detect phishing attempts, protecting users from social engineering attacks [21]. Moreover, LLMs can assist in incident response by automating tasks such as log analysis and evidence collection [16].

However, the potential of LLMs extends beyond defensive applications. Recent research, as exemplified by Happe and Cito [2], has begun to explore the use of LLMs as "sparring partners" for penetration testers, assisting in both high-level task planning and low-level vulnerability hunting. This suggests that LLMs could be weaponized for offensive cyber operations, raising ethical concerns and the need for further research in this area. For instance, the optimization of visual cryptography, as demonstrated by Ibrahim et al. [22], highlights the potential of advanced techniques in enhancing cybersecurity measures. While LLMs offer significant capabilities, ongoing research into areas like visual cryptography can further strengthen the overall security posture.

# RESEARCH GAP

The literature review reveals a significant gap in research regarding the possible weaponization of LLMs for offensive cybersecurity operations, especially concerning cyber warfare. While there are indeed some studies on the possibility of LLMs in automating and improving different phases of penetration testing including but not limited to vulnerability discovery, exploit generation, and social engineering [2], [5], [21], fewer studies have been conducted regarding how LLMs can be fine-tuned to validate the reports of VA. State-of-the-art VA currently strongly relies on automated tools, which in turn tend to generate reports with very high false-positive rates, requiring much manual effort to validate [4]. LLMs have the potential to streamline this process by automatically analyzing and validating these reports, reducing the burden on human analysts and improving the overall efficiency and accuracy of VA. As identified by Casey and Chamberlain [1], who demonstrate the ability of ChatGPT to analyze and interpret scan results, identify vulnerabilities, and even suggest potential attack vectors.

## The Proposed Approach: Fine-Tuning LLMs to Validate VA Reports

LLMs, such as Meta's Llama 3, offer a promising solution to address the limitations of current VA tools. Their ability to process and understand natural language, coupled with their knowledge of cybersecurity concepts, can be leveraged to automate and enhance the validation of VA reports. They also can employ NLP techniques to extract relevant information from VA reports, such as the identified vulnerabilities, their severity levels, and the affected assets. Additionally, this information can then be used to cross-reference with external vulnerability databases, such as the National Vulnerability Database (NVD), to verify the accuracy of the findings and gather additional context about the vulnerabilities [2].

Since LLMs have the capabilities, and with proper fine-tuning, they can be integrated with existing VA tools and security workflows to automate the validation process as they would understand VA reports efficiently. For instance, the LLM can be used to automatically test whether vulnerabilities that appear in the reports are false positives or not [16]. Consequently, this automation can significantly reduce the manual effort required for report analysis and prioritization, allowing security teams to focus on more strategic tasks. Moreover, it could potentially create a system that can compete with actual human experts.

With fine-tuning, the LLM would pipe directly with the VA tools to validate the reports on their own with minimum interference by humans. It could be a situation where LLM fires the VA tool and gets the reports itself. It would discard the obviously false-positive alerts based on knowledge inside the LLM or rules set up by the administrator. Then it should query some vulnerability databases, including NVD, Exploit-DB.com, 0day.today, cve.mitre.org, cwe.mitre.org, or any other sources, to fetch updated information and the exploitation methods. Later on, the LLM would attempt to test the vulnerabilities either by directly accessing the targets or by sending commands to other tools. Finally, the LLM should make reports of findings and present them to the administrator. Figure 4 visualizes a simple process flow of the suggested framework that is discussed in this section.

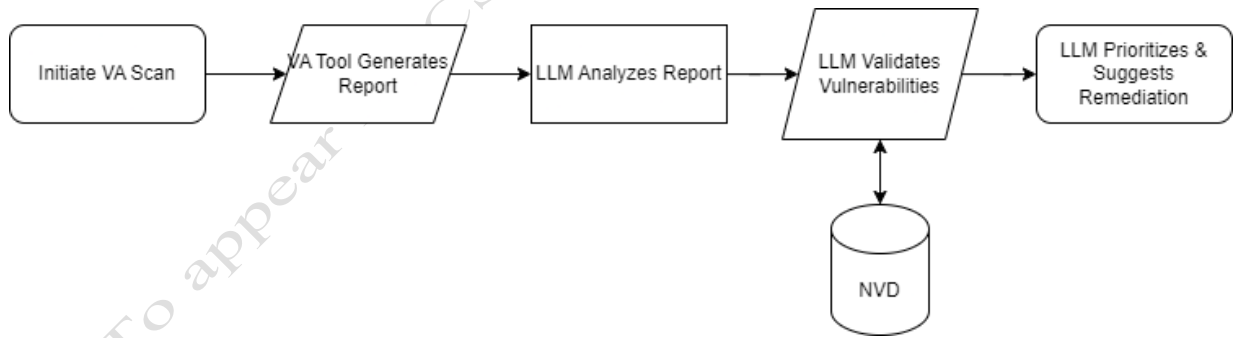

Figure 4. The suggested framework (Created by the authors)

## CONCLUSION

It has, therefore, become clear that the face of cyber war is one where LLMs take center stage with remarkable rapidity in cybersecurity defense and offense. While research on LLMs has been highly leveraged in the development of various defensive mechanisms, their utilization in the actual weaponization of offensive cyber operations has remained largely unexploited. It has also scoured through literature showing how different LLMs could automate or improve nearly every stage of a penetration test, including vulnerability discovery, exploit generation, and social engineering.

On the other hand, the literature shows that the use of LLMs for VA report validation represents a huge gap in research. This is especially opportune, considering how current tools for VA have various drawbacks and tend to provide reports with high numbers of false positives, hence requiring manual validation that can be time-consuming. The proposed research will try to fill this gap through the development and evaluation of an LLM-based model for automatic validation of reports of VA.

The potential impact of this research is substantial. The proposed LLM model automates the validation will significantly reduce the load from human analysts, improve the accuracy and efficiency of VA, and ultimately enhance the overall security posture of organizations. This research contributes to the budding literature on the offensive capabilities of LLMs, opening space for other research into the prospects of LLMs in cyber war. However, it is crucial to acknowledge and address the ethical implications of weaponizing LLMs, ensuring that their use adheres to responsible and ethical guidelines.

## FUTURE WORK

To be able to fully understand the ability to use LLMs for such potential work, there needs to be extra work done in this area. First of all, benchmarking is crucially required to identify the correct level of accuracy for LLMs. Secondly, a clear method on how to test and compare human accuracy with LLMs is a very important part as it will set a clear understanding of whether LLMs can perform better, same, or worse than human experts. Finally, since this area needs high responsibility as it could lead to huge losses [7], studies have to be done on what level of accuracy requires how much and what extra manual work to be done.

## ACKNOWLEDGMENT

In preparing this paper, I was in contact with many people, researchers, academicians, and practitioners. They have contributed towards my understanding and thoughts. Without their continued support and interest, this paper would not have been the same as presented here. I'm also grateful to Universiti Teknologi Malaysia for the opportunity they provided me with to be able to study for my PhD at their prestigious institute.

I am also indebted to the Saudi Arabian government and Imam Abdulrahman Bin Faisal University (IAU) for funding my PhD study. My fellow postgraduate students should also be recognized for their support. My sincere appreciation also extends to all my colleagues and others who have assisted on various occasions. Their views and tips are useful indeed. Unfortunately, it is not possible to list all of them in this limited space. I am grateful to all my family members.

## REFERENCES

1. D. Chamberlain and E. Casey, "Capture the Flag with ChatGPT: Security Testing with AI ChatBots," *International Conference on Cyber Warfare and Security*, vol. 19, no. 1, pp. 43–54, Mar. 2024, doi: 10.34190/ICCWS.19.1.2171.
2. A. Happe and J. Cito, "Getting pwn'd by AI: Penetration Testing with Large Language Models," *ESEC/FSE 2023 - Proceedings of the 31st ACM Joint Meeting European Software Engineering Conference and Symposium on the Foundations of Software Engineering*, pp. 2082–2086, Nov. 2023, doi: 10.1145/3611643.3613083.
3. J. Zhang, H. Bu, H. Wen, Y. Chen, L. Li, and H. Zhu, "When LLMs Meet Cybersecurity: A Systematic Literature Review," May 2024, [Online]. Available: https://arxiv.org/abs/2405.03644v1
4. V. Saber, D. Elsayad, A. M. Bahaa-Eldin, and Z. Fayed, "Automated Penetration Testing, A Systematic Review," *3rd International Mobile, Intelligent, and Ubiquitous Computing Conference, MIUCC 2023*, pp. 373–380, 2023, doi: 10.1109/MIUCC58832.2023.10278377.
5. M. C. Ghanem, T. M. Chen, and E. G. Nepomuceno, "Hierarchical reinforcement learning for efficient and effective automated penetration testing of large networks," *J Intell Inf Syst*, vol. 60, no. 2, pp. 281–303, Apr. 2023, doi: 10.1007/S10844-022-00738-0/FIGURES/13.
6. A. Nelson, S. Rekhi, M. Souppaya, and K. Scarfone, "Incident Response Recommendations and Considerations for Cybersecurity Risk Management: A CSF 2.0 Community Profile," *National Institute of Standards and Technology*, Apr. 03, 2024. https://csrc.nist.gov/pubs/sp/800/61/r3/ipd
7. Statista, "Cybercrime: monetary damage United States 2023," 2023. https://www.statista.com/statistics/267132/total-damage-caused-by-by-cybercrime-in-the-us/

8. S. Chaabani *et al.*, "LABORATORY ACCESS IMPLEMENTING QR CODE AUTHENTICATION USING OTP," *International Journal on Cybernetics & Informatics (IJCI*, vol. 12, no. 5, pp. 121–141, 2023, doi: 10.5121/ijci.2023.120511.
9. A. A. Aldulaijan, A. S. Almuhaidib, and S. O. Olatunji, "Survey of the Importance of Implementing Blockchain Technology to Support Supply-Chain Management for Identifying and Tracking Counterfeit Products," *7th International Symposium on Multidisciplinary Studies and Innovative Technologies, ISMSIT 2023 - Proceedings*, 2023, doi: 10.1109/ISMSIT58785.2023.10305002.
10. C. Greco, G. Fortino, B. Crispo, and K.-K. R. Choo, "AI-enabled IoT penetration testing: state-of-the-art and research challenges," *Enterp Inf Syst*, vol. 17, no. 9, Sep. 2023, doi: 10.1080/17517575.2022.2130014.
11. Packet Labs, "What is the difference between a VA scan and a pentest? | Packetlabs," *Packet Labs*, Jun. 11, 2020. https://www.packetlabs.net/posts/va-scan-difference-pentest/
12. F. Hussain, R. Hussain, S. Hassan, and E. Hossain, "Machine learning in IoT security: current solutions and future challenges," *IEEE Communications Surveys & Tutorials*, vol. 22, no. 3, pp. 1686–1721, 2020.
13. S. O. Olatunji *et al.*, "Machine Learning Based Preemptive Diagnosis of Parkinson's Disease Using Saudi Clinical Data: A Preliminary Case Study on Saudi Arabia Dataset," *2023 International Conference on Intelligent Data Science Technologies and Applications, IDSTA 2023*, pp. 1–6, 2023, doi: 10.1109/IDSTA58916.2023.10317845.
14. I. Butun, P. Osterberg, and H. Song, "Security of the Internet of Things: Vulnerabilities, Attacks, and Countermeasures," *IEEE Communications Surveys & Tutorials*, vol. 22, no. 1, pp. 616–644, 2020.
15. S. Chaudhary, A. O'Brien, and S. Xu, "Automated Post-Breach Penetration Testing through Reinforcement Learning," *2020 IEEE Conference on Communications and Network Security, CNS 2020*, Jun. 2020, doi: 10.1109/CNS48642.2020.9162301.
16. R. S. Jagamogan, S. A. Ismail, N. H. Hassan, and H. Abas, "Penetration Testing Procedure using Machine Learning," in *2022 4th International Conference on Smart Sensors and Application (ICSSA)*, Jul. 2022, pp. 58–63. doi: 10.1109/ICSSA54161.2022.9870951.
17. O. Valea and C. Oprisa, "Towards Pentesting Automation Using the Metasploit Framework," *Proceedings - 2020 IEEE 16th International Conference on Intelligent Computer Communication and Processing, ICCP 2020*, pp. 171–178, Sep. 2020, doi: 10.1109/ICCP51029.2020.9266234.
18. N. Koroniotis, N. Moustafa, B. Turnbull, F. Schiliro, P. Gauravaram, and H. Janicke, "A Deep Learning-based Penetration Testing Framework for Vulnerability Identification in Internet of Things Environments," *Proceedings - 2021 IEEE 20th International Conference on Trust, Security and Privacy in Computing and Communications, TrustCom 2021*, pp. 887–894, 2021, doi: 10.1109/TRUSTCOM53373.2021.00125.
19. K. C. Goh and N. Heffernan, "Toward Automated Penetration Testing Intelligently with Reinforcement Learning," National College of Ireland, Dublin, 2021. [Online]. Available: https://norma.ncirl.ie/5109/1/karchungoh.pdf
20. M. A. Al-Garadi, A. Mohamed, A. K. Al-Ali, X. Du, I. Ali, and M. Guizani, "A Survey of Machine and Deep Learning Methods for Internet of Things (IoT) Security," *IEEE Communications Surveys and Tutorials*, vol. 22, no. 3, pp. 1646–1685, Jul. 2020, doi: 10.1109/COMST.2020.2988293.
21. A. A. Masarweh and J. M. Al-Saraireh, "Enhancing the Penetration Testing Approach and Detecting Advanced Persistent Threat Using Machine Learning," Princess Sumaya University for Technology, Amman, 2021. [Online]. Available: https://www.proquest.com/docview/2682726122?pq-origsite=gscholar&fromopenview=true&sourcetype=Dissertations%20&%20Theses
22. D. Ibrahim, R. Sihwail, K. A. Z. Arrifin, A. Abuthawabeh, and M. Mizher, "A Novel Color Visual Cryptography Approach Based on Harris Hawks Optimization Algorithm," *Symmetry 2023, Vol. 15, Page 1305*, vol. 15, no. 7, p. 1305, Jun. 2023, doi: 10.3390/SYM15071305.